\documentclass[prl,aps,twocolumn,superscriptaddress]{revtex4-1}
\usepackage{amsfonts}
\usepackage{amsmath}
\usepackage{dcolumn}
\usepackage{bm}
\usepackage{tikz}
\usepackage[colorlinks=true,citecolor=blue,urlcolor=blue]{hyperref}
\usepackage{amsmath,amsfonts,amssymb,times,natbib}
\usepackage[standard]{ntheorem}

\newcommand\rket[1]{|#1\rangle}

\begin{document}

\title{High-fidelity single-shot readout of single electron spin in diamond with spin-to-charge conversion}

\author
{Qi Zhang$^{1,2,3\ast}$,
Yuhang Guo$^{1,2,3\ast}$,
Wentao Ji$^{1,2,3\ast}$,
Mengqi Wang$^{1,2,3}$,
Jun Yin$^{1,2,3}$,
Fei Kong$^{1,2,3}$,
Yiheng Lin$^{1,2,3}$,
Chunming Yin$^{1,2,3}$,
Fazhan Shi$^{1,2,3}$,
Ya Wang$^{1,2,3\dag}$,
Jiangfeng Du $^{1,2,3\dag}$
\\
\normalsize{$^{1}$ Hefei National Laboratory for Physical Sciences at the Microscale and Department of Modern Physics, University of Science and Technology of China, Hefei 230026, China.}\\
\normalsize{$^{2}$ CAS Key Laboratory of Microscale Magnetic Resonance, University of Science and Technology of China, Hefei 230026, China.}\\
\normalsize{$^{3}$ Synergetic Innovation Center of Quantum Information and Quantum Physics,}
\normalsize{University of Science and Technology of China, Hefei 230026, China.}\\
\normalsize{$^{\ast}$ These authors contributed equally to this work.}\\
\normalsize{$^\dag$ E-mail: ywustc@ustc.edu.cn, djf@ustc.edu.cn }
}

\begin{abstract}
High fidelity single-shot readout of qubits is a crucial component for fault-tolerant quantum computing~\cite{Brun2019,Camp2017} and scalable quantum networks~\cite{Wehn2018,Child2006}. In recent years, the nitrogen-vacancy (NV) center in diamond has risen as a leading platform for the above applications~\cite{Wald2014,Toga2010,Yang2016,Awsc2018,Bern2013,Pfaf2014,Hump2018,Rong2015,Abobeih2019}. The current single-shot readout of the NV electron spin relies on resonance fluorescence method at cryogenic temperature ~\cite{Robl2011}. However, the spin-flip process interrupts the optical cycling transition, therefore, limits the readout fidelity. Here, we introduce a spin-to-charge conversion method assisted by near-infrared (NIR) light to suppress the spin-flip error.  This method leverages high spin-selectivity of cryogenic resonance excitation and high flexibility of photoionization. We achieve an overall fidelity $>$ 95\% for the single-shot readout of an NV center electron spin in the presence of high strain and fast spin-flip process. With further improvements, this technique has the potential to achieve spin readout fidelity exceeding the fault-tolerant threshold, and may also find applications on integrated optoelectronic devices.
\end{abstract}
\maketitle

Resonance fluorescence method has become a commonly used method to achieve the single-shot readout of various solid-state spins such as quantum dot~\cite{Vami2010,Delt2014}, rare-earth ions in crystals~\cite{Raha2020,Kind2020}, silicon-vacancy center~\cite{Evan2018,Bhas2020} and NV center~\cite{Robl2011} in diamond. Under spin-selective excitation of optical cycling transition, the spin state is inferred according to collected spin-dependent fluorescence photon counts. However, the accompanying spin non-conservation processes usually limit the optical readout window for photon collection and induce the spin state flip error. This effect has become a significant obstacle for achieving high fidelity single-shot readout, in particular, to exceed the fault-tolerant threshold.

\begin{figure*}
	\centering
	{\includegraphics[width=1.9\columnwidth]{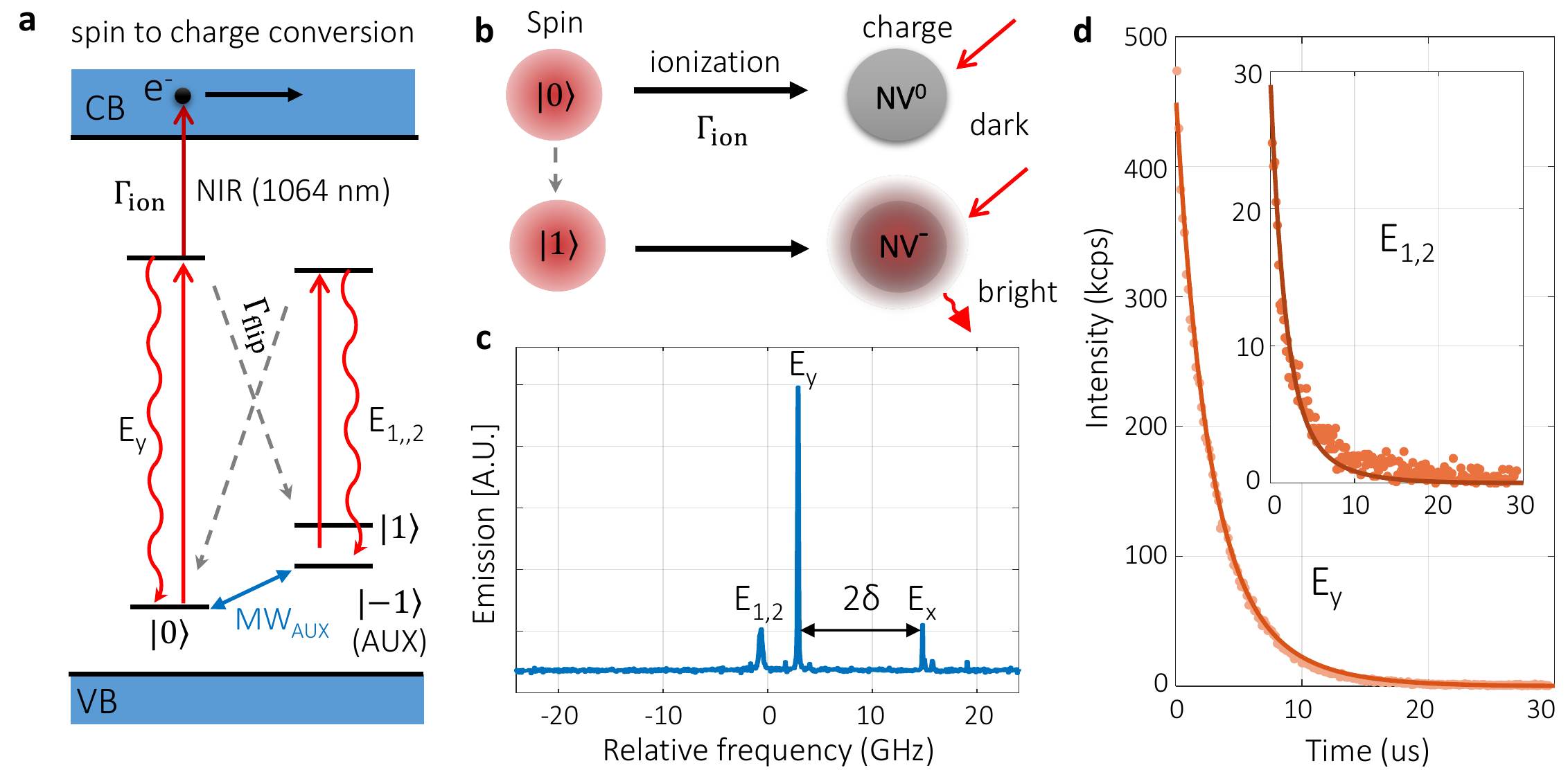}}
	\caption{Single shot readout scheme based on SCC. \textbf{(a)} Energy levels used to achieve SCC. Qubit is encoded in the ground state $\rket{0}$ and $\rket{1}$, and the
     $\rket{-1}=\rket{\rm AUX}$ state acts as the auxiliary level. The magnetic field of 585 G lifts the degeneracy between $\rket{-1}$ and $\rket{+1}$. The coherent manipulation between $\rket{0}$ and $\rket{\pm 1}$ can be realized by resonant microwave, labeled by blue arrows. E$_y$ (E$_{1,2}$) corresponds to the optical transition of the $m_S$ = 0 ($m_S$ = $\pm 1$) state. The counts rate is proportional to the excited state emission rate and the fluorescence photon collection efficiency. The key part of SCC is to ionize (dark red arrow) the excited states of $m_S = 0 $ before it substantially relaxes to the ground $\rket{\pm 1}$ states through the spin-flip relaxation process (grey dashed arrow). $\Gamma_{\rm ion}$ denotes the ionize rate, and $\Gamma_{\rm flip}$ denotes the spin-flip rate from the excited state $\rket{E_y}$ to the ground $\rket{\pm 1}$ states. A more detailed model is in the supplementary information (SI). \textbf{(b)} A schematic diagram of SCC readout. Under the illumination of 637 nm laser, NV$^{-}$ keeps fluorescing stably for a long time, while NV$^{0}$ is not excited. \textbf{(c)}  The excitation spectrum of the NV center used here at cryogenic temperature of 8K. Frequency is given relative to 470.4675 THz (637.2225 nm). The non-axial strain ($\delta$) induces a splitting of $2\delta$ = 11.8 GHz between E$_y$ and E$_x$ transitions \cite{Toga2010}.  \textbf{(d)}  Spin-flip process induces the photoluminescence (PL) decay under E$_y$ excitation (5.7 nW, saturation power $\sim$ 13 nW) with NV initially prepared in $\rket{0}$. At the final equilibrium of PL decay curve, the NV spin is pumped into $\rket{\pm 1}$. The solid line is the simulation according to the model described in SI, with the best-fitted spin-flip rate $\Gamma_{\rm flip}$ = 0.75 $\pm$ 0.02 MHz. Inset: PL decay for NV initialized to $\rket{\pm 1}$ under E$_{1,2}$ excitation (4.2 nW, saturation power $\sim$ 34 nW). From the PL decay curves, the spin initialization fidelity is estimated to be 99.7 $\pm$ 0.1 \% for $\rket{\pm 1}$ subspace and 99.8 $\pm$ 0.1 \% for $\rket{0}$ (SI).
	}\label{fig:1}
\end{figure*}

\begin{figure*}
	\centering
	{\includegraphics[width=1.9\columnwidth]{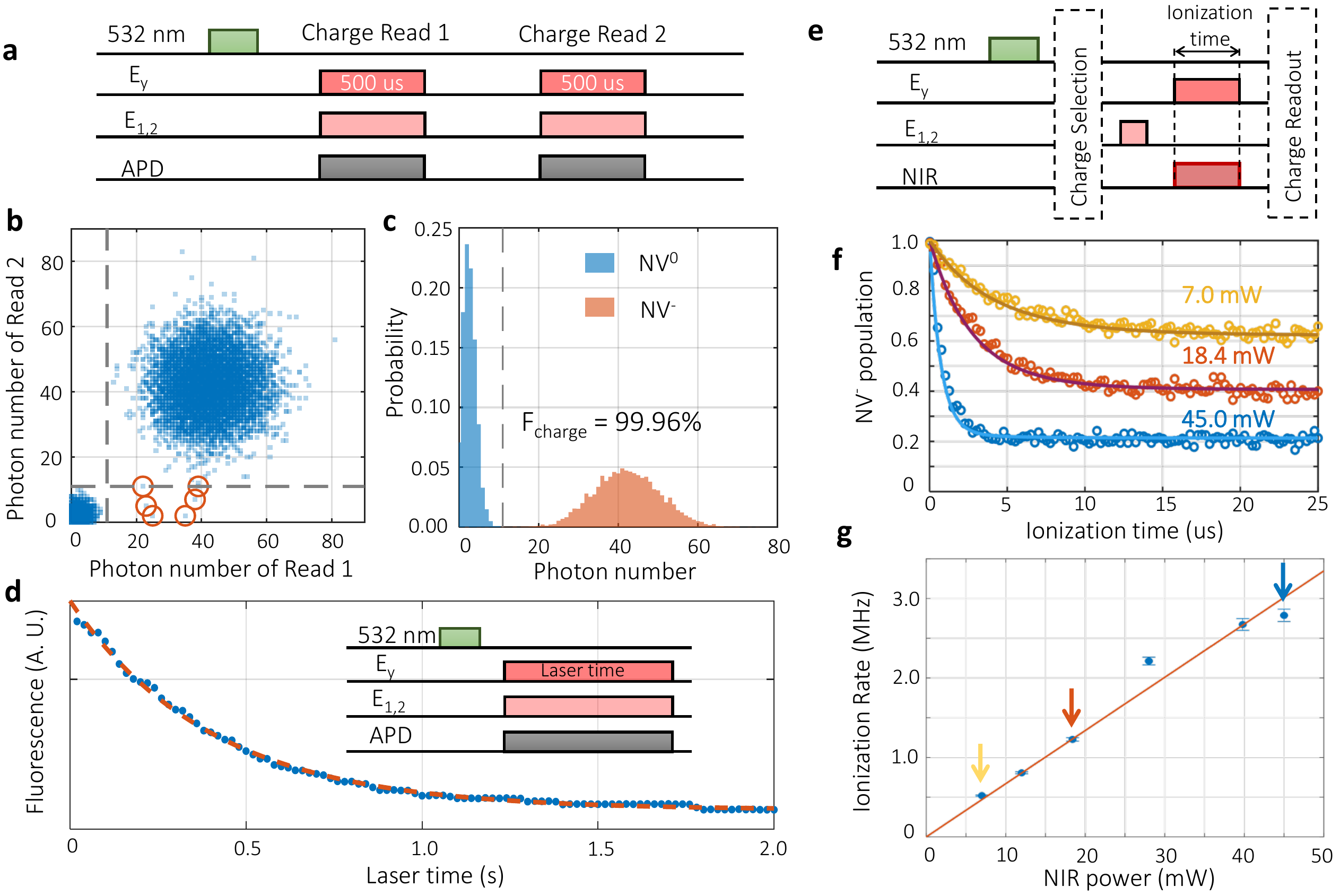}}
	\caption{
		 Non-demolition readout of charge state and ionization rate of NIR light. \textbf{(a)} Pulse sequence for the charge readout fidelity evaluation. A 3 $\mu$s pulse of 532 nm
    laser reset the population of NV$^{-}$ to be ~78\%, according the results in b and c. Both of the two charge readings use an integration window of 500 $\mu$s. \textbf{(b)} The correlation between the two consecutive charge readouts. The orange circles mark the cases with anti-correlation, all of which are NV$^{-}$ for the first readout and NV$^{0}$ for the second. \textbf{(c)} The photon number distribution of NV$^{0}$ and NV$^{-}$. \textbf{(d)} The lifetime of the charge state of NV- under E$_y$ + E$_{1,2}$ (6 + 5 nW) illumination, 400.7 $\pm$ 9.7 ms. \textbf{(e)} Pulse sequence for measuring the ionization rate under simultaneous illumination of E$_y$ and NIR light. \textbf{(f)} The ionization curves of NV$^-$ at different powers of 1064 nm. The solid lines are simulations based on different ionization rates. \textbf{(g)} The dependence of the NIR ionization rate on its power. The solid line is a linear fit to the data points, with a coefficient of 67.0 $\pm$ 6.7 kHz/mW.
	}\label{fig:2}		
\end{figure*}

\begin{figure*}
	\centering
	{\includegraphics[width=1.9\columnwidth]{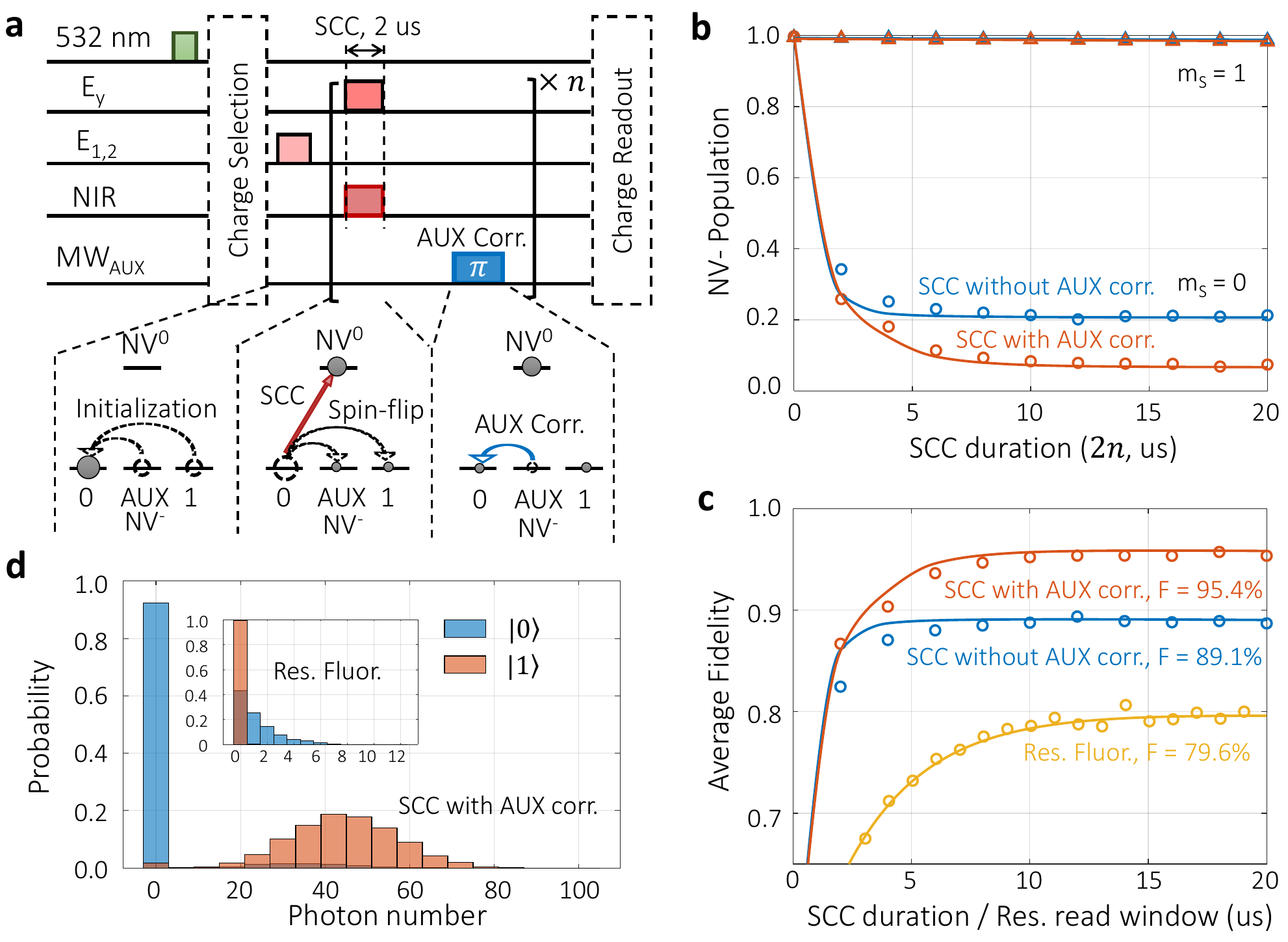}}
	\caption{
		Single-shot readout of NV electron spin state via SCC. \textbf{(a)} Pulse sequence and diagram illustrating NV spin and charge dynamics for $\rket{0}$ readout fidelity evaluation. As the spin-flip process traps some populations in  $\rket{1}$  and AUX state, an MW$_{\rm AUX}$ $\pi$ pulse `rescues' the part in AUX state back to $\rket{0}$ so that they can be ionized in the next round. The SCC pulse and ‘AUX correction’ pulse are repeated for \emph{n} rounds to get the optimal ionization. The sequence for evaluating $\rket{1}$ readout fidelity only differs in the spin initialization part, which is an E$_y$ + MW$_{\rm AUX}$ pulse of 200 $\mu$s. \textbf{(b)} NV$^{-}$  population dependence on SCC duration (2 $\mu$s $\times$ \emph{n}). The solid lines for $\rket{0}$ are simulations. The solid lines for $\rket{1}$ are linear fits to the data. \textbf{(c)} Average fidelity dependence on E$_y$ illumination time for different readout methods. In SCC methods E$_y$ illumination time equals the SCC duration, and in resonance fluorescence method it equals the read window. Blue and orange solid lines are the average of the corresponding lines in \textbf{b}. The yellow line is an exponential fit to the results of the resonance fluorescence method. \textbf{(d)} Photon number distribution of the charge readout with the SCC method. This is obtained from 20,000 measurement repetitions with NV spin initially prepared in the $\rket{0}$ (blue) and $\rket{1}$ (orange). Inset: photon distribution for the resonance fluorescence method.
	}\label{fig:3}		
\end{figure*}

A powerful method to suppress this effect is to explore optical structures for the emitters. The microstructure, such as a solid-state immersion lens, is widely used to enhance the fluorescence collection efficiency \cite{Pfaf2014,Hump2018,Robl2011,Vaha2003}. High-quality nano-cavities strongly coupled to these quantum emitters could even enhance the photon emission rate by orders of magnitude~\cite{Raha2020,Kind2020,Evan2018,Bhas2020}. Despite these significant achievements, the practical application of such a high-quality cavity remains technically challenging. Extensive engineering works are required to obtain the high-quality cavity, place the emitter into the optimal cavity position, and tune the frequency on-demand. Besides, the fabrication process introduces unwanted strain and surface defects~\cite{Lonc2013}, which may degrade the spin and optical properties~\cite{Robl2011}.

Here, we demonstrate a new method to achieve a single-shot readout of NV center electron spin by combing a spin-selective photoionization process. The spin state is on-demand converted into charge state before the spin-flip relaxation becomes significant (Fig.~\ref{fig:1}a,b). Then the charge state is measured with near unity fidelity thanks to their stability under optical illumination. The essence of this approach is to enhance the ratio of ionization rate ($\Gamma_{\rm ion}$) to the spin-flip rate ($\Gamma_{\rm flip}$).

The experiments are performed on a bulk NV center inside a solid immersion lens at a cryogenic temperature of 8 K. The measurement scheme utilizes the cycling transition E$_y$ that connects excited and ground states with spin projection $m_S=0$ (Fig.~\ref{fig:1}a), and the E$_{1,2}$ transition connecting states with spin projection $m_S=\pm 1$. The corresponding optical transitions is shown in Fig.~\ref{fig:1}c. The fabrication of the solid immersion lens introduced non-axial strain $\delta$ = 5.9 GHz to the NV center used. Therefore, a spin-flip rate $\Gamma_{\rm flip}$ of 0.75 $\pm $ 0.02 MHz is observed (Fig.~\ref{fig:1}d), much faster than previous reports with low strains~\cite{Robl2011}. Under selective excitation of E$_y$, spin state $\rket{0}$ could be pumped to the excited state, and be further ionized to charge state NV$^0$ under another NIR laser excitation (1064 nm, Fig.~\ref{fig:1}a). In contrast, $\rket{\pm 1}$ will not be excited and stay at charge state NV$^-$. Such a deterministic SCC differs from previous work using non-resonant excitation to enhance the readout efficiency of NV center~\cite{Shie2015,Hopp2016,Jask2019,Hopp2018,Jaya2018,Hopp2020}.

To verify the photoionization process, we first characterize the charge state readout. Under simultaneous excitation of E$_y$ and E$_{1,2}$ transitions, NV$^-$ emits photons regardless of the spin state while leaving NV$^0$ in the unexcited dark state. The charge state can thus be determined from the detected photon number during the integration window. We evaluate the charge readout fidelity by measuring the correlation between two consecutive readouts (Fig.~\ref{fig:2}a). The correlation results with an integration window of 500 $\mu$s is shown in Fig.~\ref{fig:2}b and the statistical distribution of the photon number is shown in Fig.~\ref{fig:2}c. As expected, the NV$^-$ state is distinguishable from the NV$^0$ state according to the photon counts(Fig.~\ref{fig:2}c). More importantly, a strong positive correlation is observed, except for six anti-correlation cases. And all these anti-correlation cases (circles in Fig.~\ref{fig:2}b) comes from initial NV$^-$ transforming to NV$^0$. This indicates a unity readout fidelity for NV$^0$ state and 99.92 $\pm$ 0.03 \% readout fidelity for NV$^-$ state. To understand the tiny readout imperfection for NV$^-$ state, we measure its lifetime under the continuous optical readout sequence. As shown in Fig.~\ref{fig:2}d, one observes a lifetime of 400.7 $\pm$ 9.7 ms for NV$^-$ state, which causes a charge conversion error of 0.12\% during the charge state readout, comparable to the observed imperfection. The average non-demolition charge readout fidelity is 99.96 $\pm$ 0.02 \%.

\begin{figure*}
	\centering
	{\includegraphics[width=1.4\columnwidth]{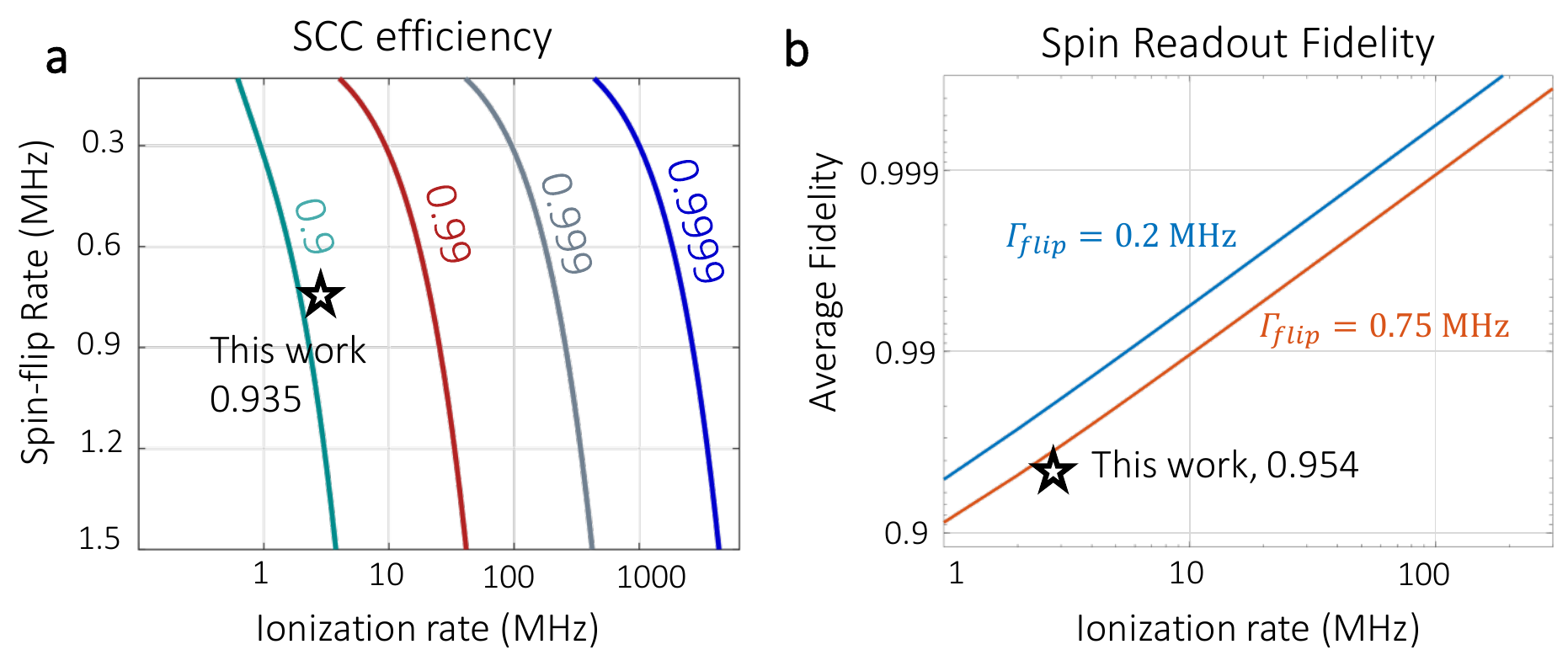}}
	\caption{
		\text{Exceeding the 99.9\% fault-tolerant threshold.}
		\textbf{(a)} Effects of ionization rate and spin-flip rate on SCC efficiency, which is indicated by the number next to each curve. \textbf{(b)} Dependence of overall spin readout fidelity on ionization rate at two different spin-flip rates. The orange line corresponds to spin-flip rate observed in this work. The blue line is a prediction for an NV center with a low spin-slip rate reported in Ref.~\cite{Robl2011}.
	}\label{fig:4}
\end{figure*}

With the non-demolition charge readout, we investigate the ionization by various NIR illumination. We first initialize the charge state to NV$^-$ by a 532 nm laser pulse and measurement-based charge state post-selection. Then a 20 $\mu$s pulse of E$_{1,2}$ initializes the spin to state $\rket{0}$. After the charge and spin initialization, the SCC process is applied, followed by a charge state readout (Fig.~\ref{fig:2}e). In contrast to the long charge lifetime of 400.7 ms observed in the absence of NIR laser (Fig.~\ref{fig:2}d), the NV$^-$ population decays fast on the timescale of microseconds after simultaneous illumination of E$_y$ and NIR light (Fig.~\ref{fig:2}f). However, the NV$^-$ population saturation level does not reach at 0, indicating that in some cases $\rket{0}$ goes through the spin-flip process and gets trapped in $\rket{\pm 1}$, which does not ionize. As the NIR power increases, the NV$^-$ population decay faster and saturates at lower levels.
To estimate the ionization rate $\Gamma_{\rm ion}$, we develop an extensive model including a more complicated energy structure as described in SI. The model uses independently measured quantities with only one free parameter $\Gamma_{\rm ion}$, which can closely match the data (Fig.~\ref{fig:2}f). The extracted ionization rate is proportional to the NIR laser power (Fig.~\ref{fig:2}g). This indicates that the NV center is most likely to be ionized from the excited state by absorbing a single 1064 nm photon. The obtained coefficient of 67.0$\pm$6.7 kHz/mW is much lower than the 1.2$\pm$0.33 MHz/mW previously estimated at room temperature~\cite{Meir2018}, which requires further study in the future.

The highest $\Gamma_{\rm ion}$ obtained is 2.79$\pm$0.08 MHz (for 45.0 mW after the objective), only 3.7 times of $\Gamma_{\rm flip}$ = 0.75$\pm$0.02 MHz. One limitation is the output power of current CW NIR laser. The other is the high loss of laser power density on NV center due to transmission reduction and chromatic aberration of the objective. The resulting single-shot fidelity is 89.1 $\pm$ 0.2 \% (blue line in Fig.~\ref{fig:3}c). To improve the conversion efficiency ($\rket{0} \rightarrow$ NV$^0$) under current conditions, we consider a correction scheme by utilizing the auxiliary level $m_S$ = -1. As shown in Fig.~\ref{fig:3}a, the leakage population from $\rket{0}$ to the AUX state, is transferred back to $\rket{0}$ state through an MW$_{\rm AUX}$ $\pi$ pulse. With this correction, the $\rket{0}$ is converted into NV$^0$ with higher efficiency, while conversion of state $\rket{1}$ is not affected (Fig.~\ref{fig:3}b). The resulting single shot fidelity is shown in Fig.~\ref{fig:3}c. With about 10 $\mu$s SCC duration, the average fidelity reaches its maximum of F$_{\rm avg}$ = 1/2 (F$_{\rket{0}}$ + F$_{\rket{1}}$) = 95.4 $\pm$ 0.2 \%. The corresponding histogram is given in Fig.~\ref{fig:3}d.
We also compare the SCC method with the resonance fluorescence method for the single-shot readout. Due to the sizeable spin-flip rate, the optimal average fidelity with resonance fluorescence method is 79.6 $\pm$ 0.8 \% (Fig.~\ref{fig:3}c,d), much lower than previous reports with low-strain NV centers~\cite{Bern2013,Pfaf2014,Hump2018,Robl2011}.

The main limiting factor for our single-shot readout fidelity is the SCC efficiency. It depends on both the ionization rate and the spin-flip rate. Fig.~\ref{fig:4}a shows the simulation results using our model(SI). The larger ratio $\Gamma_{\rm ion}/\Gamma_{\rm flip}$ is, the higher efficiency could be achieved. In practice, $\Gamma_{\rm flip}$ has a lower bound solely determined by the intrinsic property of NV center. In contrast, $\Gamma_{\rm ion}$ is convenient to increase by using high power NIR laser and good transmission objective. For a lower $\Gamma_{\rm flip} \sim$ 0.2 MHz~\cite{Robl2011}, a modest NIR power $>$ 1 W on the diamond could achieve an average single-shot readout fidelity exceeding 99.9\% (Fig.~\ref{fig:4}b), meeting the requirement for fault-tolerant quantum computing and networks~\cite{Camp2017,Mura2014}.

In summary, we demonstrate a NIR-assisted SCC method for the singe-shot readout of electron spin with fidelity of 95.4\%. Different from previous methods which requires careful engineering to improve the emission rate and photon collection efficiency, our method only need an additional NIR beam. By directly controlling the NIR power, the above calculations suggest that the NIR-assisted SCC is an experimentally feasible approach towards spin readout exceeding the fault-tolerant threshold.

Another promising application of single-shot SCC is high-efficiency quantum sensing as discussed in a recent work~\cite{Irbe2020}. Because most of the bio-molecules are rarely affected by the NIR light, the NIR-assisted SCC demonstrated here is helpful to avoid photo-damage on the bio-samples~\cite{Shi2015,Shi2018,Cheng2018}. The SCC scheme also has the potential for applications on integrated quantum devices~\cite{Bren2015,Bour2015,Gulk2016,Hrub2017,Siyu2019}. At present, the photoelectric detection of single NV centers relies on measuring photocurrent from multiple ionizations~\cite{Siyu2019}. The deterministic SCC opens the possibility for achieving optoelectronic single-shot readout of solid spins, potentially utilizing the single-electron transistor as charge reading head~\cite{Yin2013,Zhan2019}. Finally, since SCC readout is a demolition method for electron spins, nuclear spins weakly coupled to the NV center would allow the projective readout~\cite{Jiang2009,Neumann2010,Maur2012}.

\emph{Notice.-}
There is another similar work using visible laser to achieve single-shot SCC under severe conditions~\cite{Irbe2020}.

\emph{Acknowledgement.-}
The authors are grateful to Sven Rogge, Milos Nesladek, Friedemann Reinhard, and Guanglei Cheng for helpful discussions.
This work is supported by the National Key R$\&$D Program of China (Grant No. 2018YFA0306600, 2017YFA0305000, 2016YFA0502400),
the NNSFC (Grants No. 11775209, 81788101,11761131011, 11722544),
the CAS (Grants No. GJJSTD20170001, No. QYZDY-SSW-SLH004),
Anhui Initiative in Quantum Information Technologies (Grant No. AHY050000),
the Fundamental Research Funds for the Central Universities.



\end{document}


\title{Supplementary Information for High-fidelity single-shot readout of single electron spin in diamond with spin-to-charge conversion}

\author
{Qi Zhang$^{1,2,3\ast}$,
Yuhang Guo$^{1,2,3\ast}$,
Wentao Ji$^{1,2,3\ast}$,
Mengqi Wang$^{1,2,3}$,
Jun Yin$^{1,2,3}$,
Fei Kong$^{1,2,3}$,
Yiheng Lin$^{1,2,3}$,
Chunming Yin$^{1,2,3}$,
Fazhan Shi$^{1,2,3}$,
Ya Wang$^{1,2,3\dag}$,
Jiangfeng Du $^{1,2,3\dag}$
\\
\normalsize{$^{1}$ Hefei National Laboratory for Physical Sciences at the Microscale and Department of Modern Physics, University of Science and Technology of China, Hefei 230026, China.}\\
\normalsize{$^{2}$ CAS Key Laboratory of Microscale Magnetic Resonance, University of Science and Technology of China, Hefei 230026, China.}\\
\normalsize{$^{3}$ Synergetic Innovation Center of Quantum Information and Quantum Physics,}
\normalsize{University of Science and Technology of China, Hefei 230026, China.}\\
\normalsize{$^{\ast}$ These authors contributed equally to this work.}\\
\normalsize{$^\dag$ E-mail: ywustc@ustc.edu.cn, djf@ustc.edu.cn }
}
\maketitle

\section{Experimental apparatus}
All the experiments in this work are performed on a home-built low-temperature ODMR setup. The sample is hosted at the temperature of 8 K in a closed-cycle optical cryostat (Montana Instruments Cryostation S200). The sample is positioned by a XYZ piezo stack (Attocube), at the focal point of a 0.9 NA objective (Olympus MPLFLN100x). The objective is mounted on the side wall of the vacuum chamber. A feedback loop is used to stabilize the temperature of the objective at room temperature. We use three lasers to excite the NV center: a traditional 532 nm laser (Changchun New Industries Optoelectronics Technology) to reset charge state and two tunable 637 nm lasers (New focus TLB-6704-OI, Toptica DLC DL PRO HP 637) to perform E$_y$ and E$_{1,2}$ resonant excitation respectively. The wavelengths of the 637 nm lasers are stabilized using a wavemeter (HighFiness WSU-10). We use a 1064 nm laser (Changchun New Industries Optoelectronics Technology MSL-III-1064) to ionize the NV center. A permanent magnet is positioned near the sample by another set of XYZ piezo stack (Attocube) to produce a static magnetic field. Microwave pulses are generated by an arbitrary waveform generator (Keysight M8190A).  After amplification (minicircuit), the Microwave pulses are fed to a gold strip line fabricated on top of the sample.

\section{Spin state initialization fidelity}

To estimate the spin initialization fidelity, we prepare $\rket{0}$ ($\rket{\pm1}$) state by a E$_{1,2}$ (E$_y$) pulse and record the resonance fluorescence time trace under E$_y$ (E$_{1,2}$) illumination. We fit the results with double exponential decay curves and extract the initial and final equilibrium count rate. The initial count rate for $\rket{0}$ ($\rket{\pm1}$) state is 166.7 kctps (39.4 kctps). Note that final count rate of the $\rket{\pm1}$ result is higher than the $\rket{0}$ result. This is due to a higher background count rate from the E$_{1,2}$ laser (2.8 kctps) than from the E$_y$ laser (0.5 kctps). After subtracting the background, we can estimate a remaining population of 0.17$\pm$0.06 \% in $\rket{0}$  state after E$_y$ pumping, and 0.06$\pm$0.15 \% in $\rket{\pm1}$ state after E$_{1,2}$ pumping, giving state initialization fidelity of 99.83$\pm$0.06 \% for $\rket{\pm1}$ state and 99.94$\pm$0.15 \% for $\rket{0}$ state. In the main text, to initialize to $\rket{0}$ state, we apply a 20 $\mu$s E$_{1,2}$ pulse, corresponding to a 99.82\% initialization fidelity. To initialize to $\rket{+1}$ state, along with the E$_y$ laser pumping $\rket{0}$ state, we apply an additional microwave pulse to flip the population from $\rket{-1}$ back to $\rket{0}$ state. After 200 $\mu$s E$_y$ and microwave pulse, we can estimate the remaining population in both $\rket{0}$ and $\rket{-1}$ as 0.17\%, giving 99.66\% initialization fidelity into $\rket{+1}$ state.

\begin{figure}[htp]
	\centering
	{\includegraphics[width=0.6\columnwidth]{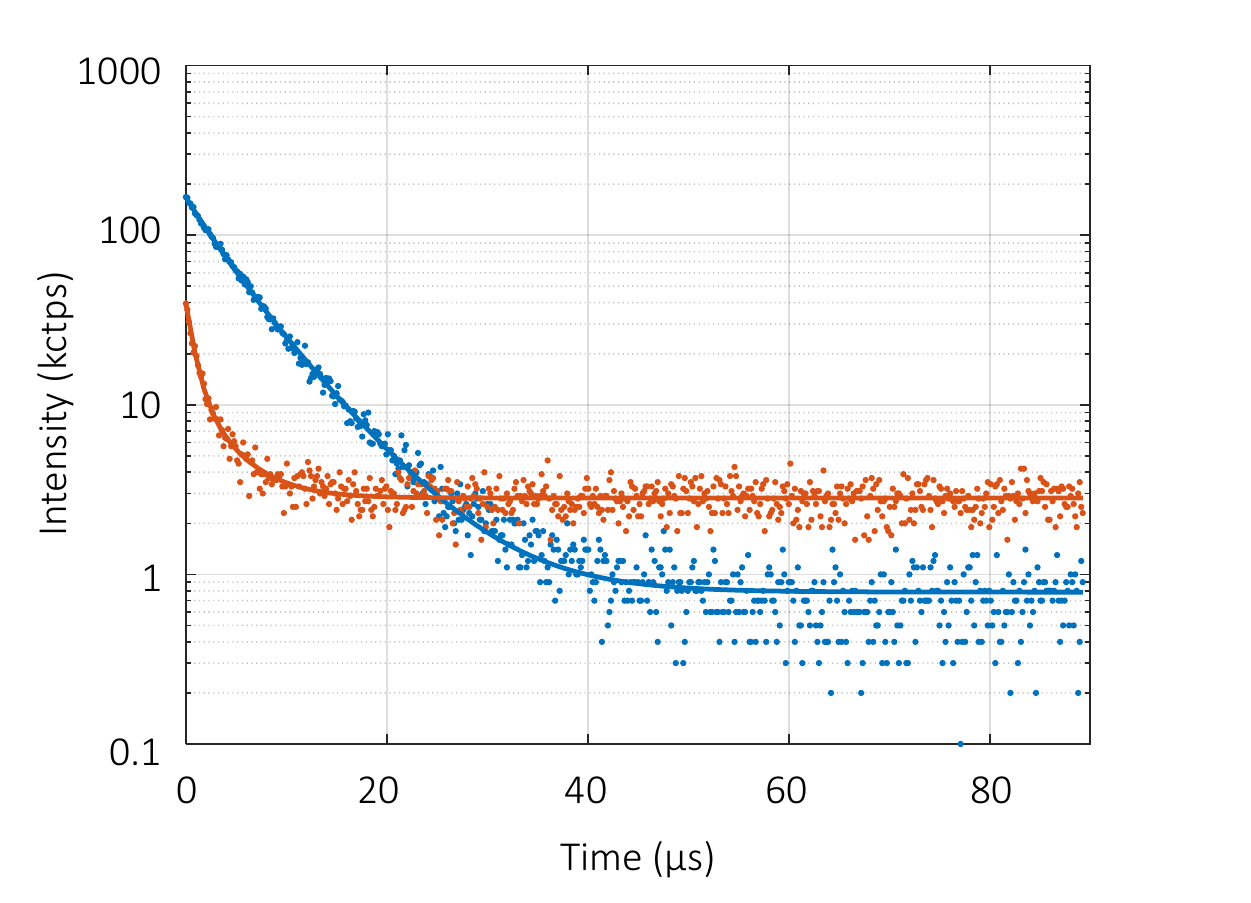}}
	\caption{Photoluminescence decay of NV center, initially prepared in $\rket{0}$ (blue) and $\rket{\pm1}$ (orange).}
\end{figure}

\section{Improving charge readout fidelity }

\begin{figure}[htp]
	\centering
	{\includegraphics[width=0.6\columnwidth]{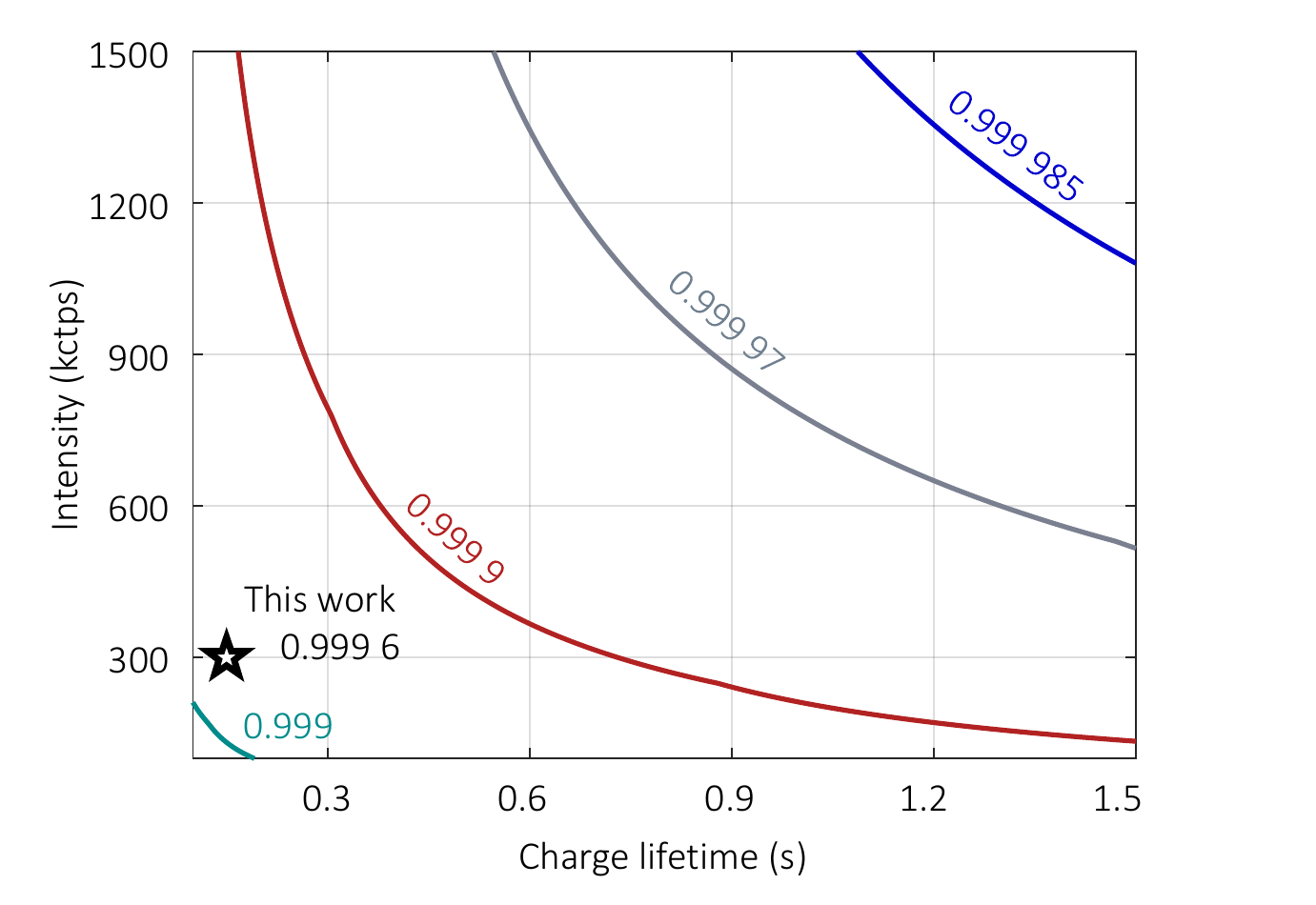}}
	\caption{Effect of photon count rate and charge lifetime on charge readout fidelity.}
\end{figure}

The charge readout fidelity mainly depends on the photon count rate and charge lifetime. Resonance fluorescence count of NV$^{-}$ is proportional to spontaneous emission rate and fluorescence collection efficiency, while NV$^{0}$ count is mainly affected by background fluorescence. Here we assume a background fluorescence of 2.5 kctps, and optimize the readout window by balancing the effects of photon shot noise and charge state lifetime, to obtain the best non-demolition charge readout fidelity. The black star marked our current level of charge readout (fidelity 99.96\%). For single-shot charge readout, higher fidelity could be achieved by prolonging the reading time.

\section{Model for spin and charge dynamics}

\begin{figure}[htp]
	\centering
	{\includegraphics[width=0.65\columnwidth]{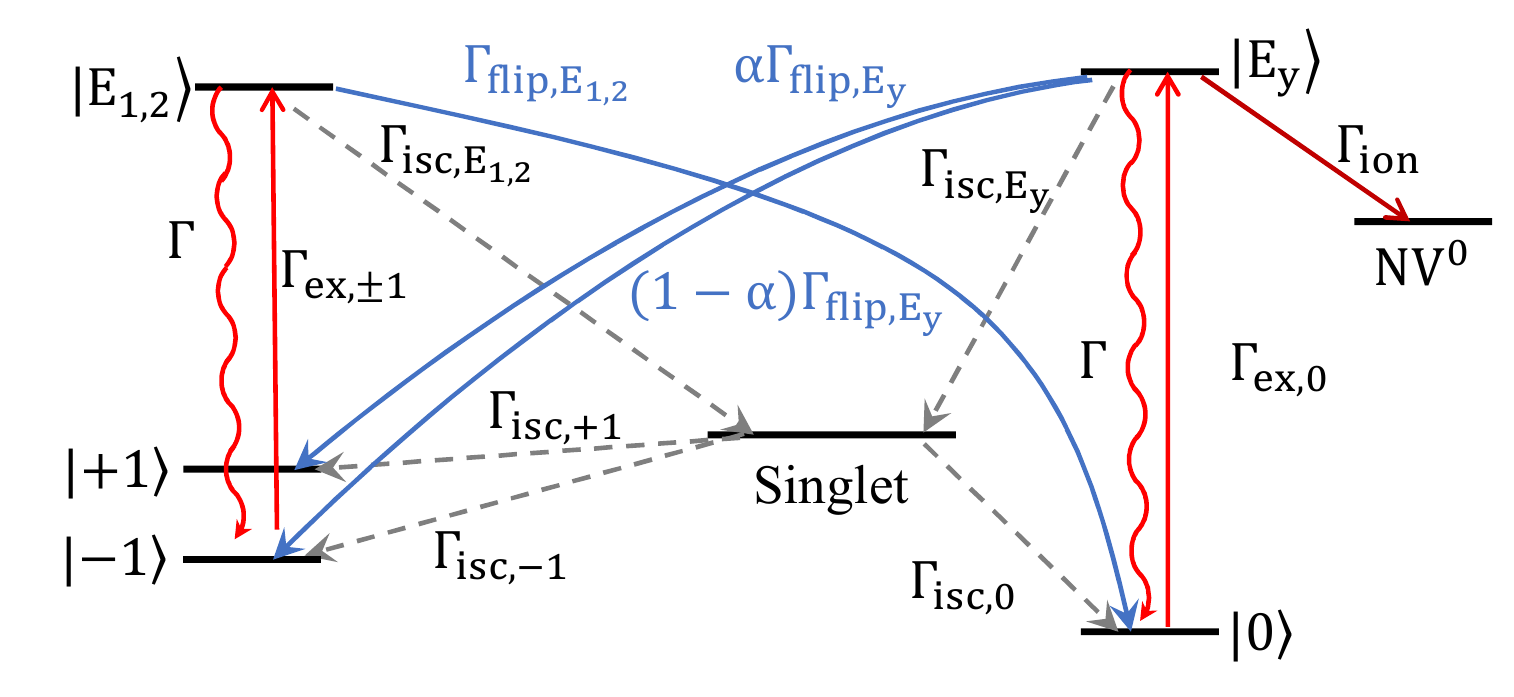}}
	\caption{
		Detailed energy level diagram used in the simulation.
		The optical transitions are denoted by red arrows.
		The ionization process is denoted by dark red arrow.
		The spin flip transitions are denoted by blue arrows.
		The inter-system crossing(ISC) transitions are denoted by dashed grey arrows.}\label{fig:S3}
\end{figure}

In order to understand the charge dynamics observed in experiment, we consider a 7 level energy diagram relevant to the SCC process, as depicted in Fig.S3. The dynamics among different levels can be described by the following rate equation
\begin{alignat*}{2}
&\frac{d}{dt}P_0 &&= -\Gamma_{ex,0}P_0 + \Gamma P_{E_y} + \Gamma_{isc,0}P_{singlet} + \Gamma_{flip,E_{1,2}}P_{E_{1,2}}\\
&\frac{d}{dt}P_{+ 1} &&= -\Gamma_{ex,\pm1}P_{+1} + 0.5\Gamma P_{E_{1,2}} + \alpha \Gamma_{flip,E_y}P_{E_y} + \Gamma_{isc,+1}P_{singlet}\\
&\frac{d}{dt}P_{- 1} &&= -\Gamma_{ex,\pm1}P_{-1} + 0.5\Gamma P_{E_{1,2}} + (1-\alpha) \Gamma_{flip,E_y}P_{E_y} + \Gamma_{isc,-1}P_{singlet}\\
&\frac{d}{dt}P_{E_y} &&= \Gamma_{ex,0}P_{0} -\Gamma P_{E_{y}} - \Gamma_{isc,E_y}P_{E_y} - \Gamma_{ion}P_{E_y} + \Gamma_{flip,E_y}P_{E_y}\\
&\frac{d}{dt}P_{E_{1,2}} &&= \Gamma_{ex,\pm1}(P_{+1}+P_{-1}) -\Gamma P_{E_{1,2}} - \Gamma_{isc,E_{1,2}}P_{E_{1,2}} - \Gamma_{flip,E_{1,2}}P_{E_{1,2}}\\
&\frac{d}{dt}P_{singlet} &&= \Gamma_{isc,E_y}P_{E_y} + \Gamma_{isc,E_{1,2}}P_{E_{1,2}} - (\Gamma_{isc,0} - \Gamma_{isc,+1} + \Gamma_{isc,-1})P_{singlet}\\
&\frac{d}{dt}P_{NV^0} &&= \Gamma_{ion}P_{E_y}
\end{alignat*}
where the spontaneous emission rate $\Gamma = 77$ MHz ~\cite{Meir2018}, the excitation rate $\Gamma_{ex,0}$ is estimated directly according to the fluorescence saturation curve,  $\Gamma_{ex,0}=\frac{\rm PL}{\rm PL_{sat}-PL}\Gamma = \frac{480}{1157-480}*77 \sim 54$ MHz. $\Gamma_{isc,+1}+\Gamma_{isc,-1}+\Gamma_{isc,0} = 1/\tau_{singlet} \sim 0.3$ MHz is the singlet decay rate at 8K \cite{Robl2011}. The branching ratio from the singlet state to the ground state is $\Gamma_{isc,+1}:\Gamma_{isc,-1}:\Gamma_{isc,0}$ = 1:1:8 according to the literature \cite{Kalb2018}. With these fixed parameters, we further determine the rates $\Gamma_{ex, 0}$, $\Gamma_{isc,E_{y}}$ and $\Gamma_{flip,E_{y}}$ by fitting the fluorescence decay under E$_y$ excitation. Noted that since $\rket{\pm 1}$ is not excited under E$_y$ illumination, the rates $\Gamma_{ex,\pm 1}$, $\Gamma_{isc,E_{1,2}}$ and $\Gamma_{flip,E_{1,2}}$ are set to zero in the PL decay simulation of E$_y$ excitation.  Similarly, the rates $\Gamma_{ex, \pm1}$, $\Gamma_{flip,E_{1,2}}$ and $\Gamma_{flip,E_{1,2}}$ are determined by fitting the fluorescence decay under E$_{1,2}$ excitation. With all these parameters derived, the ionization rates $\Gamma_{ion}$ are then obtained by fitting the charge conversion curves in Fig. 2f. The derived ionization rates are given in Table~\ref{tab1}. The branching factor $\alpha$ of the spin-flip process from $\rket{E_y}$ to $\rket{\pm1}$ (Fig.~\ref{fig:S3}) is determined by fitting the pulsed SCC curve in Fig. 3b. All the parameters derived with our model are given in Table~\ref{tab2}.

\begin{table}[!t]
	\begin{minipage}[!t]{\columnwidth}								
	\centering
	\caption{Ionization rate corresponding to Fig.~2f}
	\begin{tabular}{c@{\hspace{1cm}}c}
		\toprule
		Effective NIR laser power [mW] & $\Gamma_{\rm ion}$ [MHz] \\
		\hline
		7.0 &  0.52 $\pm$ 0.01 \\
		12.0 & 0.81 $\pm$ 0.01 \\
		18.4 & 1.23 $\pm$ 0.02 \\
		28.0 & 2.21 $\pm$ 0.05 \\
		39.8 & 2.67 $\pm$ 0.07 \\
		45.0 & 2.79 $\pm$ 0.08 \\
		\hline
	\end{tabular}\label{tab1}
	\end{minipage}		

	\begin{minipage}[!t]{\columnwidth}		
	\centering
	\caption{Parameters for the fitting}
	\begin{tabular}{c@{\hspace{1cm}}c}
		\toprule
		Parameter & Fitted value \\
		\hline
		$\Gamma_{ ex,0}$ & 54.906 $\pm$ 0.372 MHz \\
		$\Gamma_{ flip,E_y}$ & 0.752 $\pm$ 0.005 MHz \\
		$\Gamma_{ isc,E_y}$ & 0.132 $\pm$ 0.004 MHz \\
		$\Gamma_{ ex,\pm1}$ & 1.209 $\pm$ 0.449 MHz \\
		$\Gamma_{ flip,E_{1,2}}$ & 2.010 $\pm$ 4.277 MHz \\
		$\Gamma_{ isc,E_{1,2}}$ & 52.760 $\pm$ 0.376 MHz \\
		$\alpha$ & 0.275 $\pm$ 0.191\\
		\hline
	\end{tabular}\label{tab2}
	\end{minipage}			
\end{table}